\documentclass[manuscript, screen, acmlarge]{acmart}

\renewcommand\footnotetextcopyrightpermission[1]{}

\usepackage{hanging}
\usepackage[utf8]{inputenc}
\usepackage{multirow}
\usepackage{xcolor}
\usepackage[utf8]{inputenc}

\newcommand{\SysName}{MARVEL}

\AtBeginDocument{%
  }

\usepackage{fancyhdr}
\begin{document}

\title{Unified Acoustic Representations for Screening Neurological and
Respiratory Pathologies from Voice} 

\author{Ran Piao}
\affiliation{
  \institution{Eindhoven University of Technology}
  \city{Eindhoven}
  \country{The Netherlands}
}
\email{r.piao@tue.nl}

\author{Yuan Lu}
\affiliation{
  \institution{Eindhoven University of Technology}
  \city{Eindhoven}
  \country{The Netherlands}
}
\email{y.lu@tue.nl}

\author{Hareld Kemps}
\affiliation{
  \institution{Máxima MC Hospital}
  \city{Eindhoven}
  \country{The Netherlands}
}
\email{h.kemps@mmc.nl}

\author{Tong Xia}
\affiliation{
  \institution{Tsinghua University}
  \city{Beijing}
  \country{China}
}
\email{tongxia@mail.tsinghua.edu.cn}

\author{Aaqib Saeed}
\affiliation{
  \institution{Eindhoven University of Technology}
  \city{Eindhoven}
  \country{The Netherlands}
}
\email{a.saeed@tue.nl}

\renewcommand{\shortauthors}{Piao et al.}

\begin{abstract}
Voice-based health assessment offers unprecedented opportunities for scalable, non-invasive disease screening, yet existing approaches typically focus on single conditions and fail to leverage the rich, multi-faceted information embedded in speech. We present \SysName~ (Multi-task Acoustic Representations for Voice-based Health Analysis), a privacy-conscious multitask learning framework that simultaneously detects nine distinct neurological, respiratory, and voice disorders using only derived acoustic features, eliminating the need for raw audio transmission. Our dual-branch architecture employs specialized encoders with task-specific heads sharing a common acoustic backbone, enabling effective cross-condition knowledge transfer. Evaluated on the large-scale Bridge2AI-Voice v2.0 dataset, \SysName~ achieves an overall AUROC of 0.78, with exceptional performance on neurological disorders (AUROC = 0.89), particularly for Alzheimer's disease/mild cognitive impairment (AUROC = 0.97). Our framework consistently outperforms single-modal baselines by 5-19\% and surpasses state-of-the-art self-supervised models on 7 of 9 tasks, while correlation analysis reveals that the learned representations exhibit meaningful similarities with established acoustic features, indicating that the model's internal representations are consistent with clinically recognized acoustic patterns. By demonstrating that a single unified model can effectively screen for diverse conditions, this work establishes a foundation for deployable voice-based diagnostics in resource-constrained and remote healthcare settings.

\end{abstract}

\ccsdesc[500]{Computing methodologies~Multitask learning}
\ccsdesc[500]{Computing methodologies~Machine learning}
\ccsdesc[300]{Applied computing~Health care information systems}
\ccsdesc[300]{Human-centered computing~Accessibility technologies}

\keywords{Voice-based diagnostics, Multitask learning, Acoustic feature representations, Neurological and respiratory disorder detection}

\maketitle

\section{Introduction}

The human voice, intrinsically natural, non-invasive, and readily collectable, is increasingly recognized as a potent and low-cost digital biomarker for health~\citep{fagherazzi2021voice, naserclaes2021speech}. Acoustic properties embedded within speech—such as pitch, prosody, speaking rate, and loudness—extend beyond conveying linguistic and emotional states, offering valuable indicators of an individual's cognitive, neurological, respiratory, and vocal well-being \cite{luz2021detecting}. These insights underpin the potential of voice analysis as a scalable modality for the early screening and monitoring of diverse conditions, including Parkinson’s disease (PD), mild cognitive impairment (MCI), airway-related diseases, and various voice disorders~\citep{tsanas2011novel}.

The clinical manifestations of these conditions often leave discernible imprints on vocal production. For instance, respiratory ailments like airway stenosis can lead to abnormal breath control and altered vocal tract coordination \citep{gelbard2016disease}, while voice disorders such as vocal fold paralysis or structural lesions typically result in unstable vocal quality~\citep{roy2007toward}. Neurological or cognitive conditions like PD or MCI can impact prosody and fluency, often yielding slowed or monotonous speech patterns~\citep{skodda2011vocal, martinez2021ten}. Traditional diagnostic workflows frequently necessitate in-person visits to healthcare facilities or the use of cumbersome physiological sensors. In contrast, speech-based analysis presents a significantly lower-burden, remote-friendly alternative, amenable to passive or active collection during everyday interactions, thereby easing patient engagement and facilitating longitudinal monitoring \citep{robin2020evaluation}.

The proliferation of mobile devices equipped with local processing capabilities has paved the way for deploying speech-based diagnostic systems in real-world, privacy-sensitive settings. Crucially, raw speech recordings, which may contain sensitive personal information, do not need to be transmitted. Instead, acoustic representations like Mel-frequency cepstral coefficients (MFCCs) or log-Mel spectrograms can be extracted locally on-device~\citep{backstrom2023privacy}. This on-edge processing paradigm safeguards patient privacy, promotes ethical data handling, and enables the scalable deployment of voice-based health technologies across diverse healthcare ecosystems.

\begin{figure*}[t]
  \centering
  \includegraphics[width=0.9\textwidth]{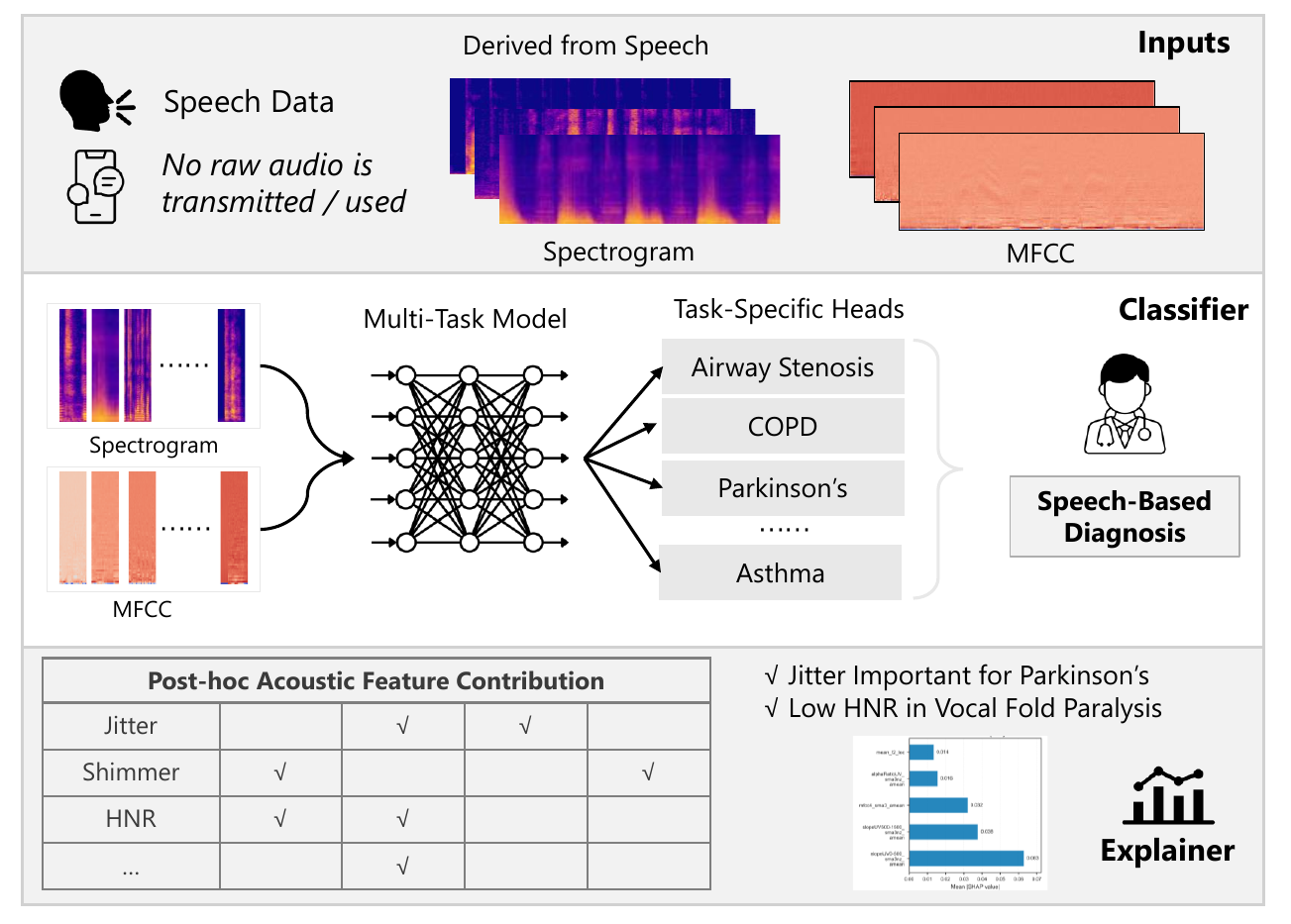}
  \caption{Overview of our proposed multi-task dual-modal \SysName~ framework.}
  \label{fig:teaser}
\end{figure*}

In this work, we introduce a multitask disorder classification framework, \SysName~ (Multi-task Acoustic Representations for Voice-based Health Analysis), that operates exclusively on such derived speech features.An overview of the our framework is shown in Figure~\ref{fig:teaser}. Our model employs a dual-branch architecture with a shared acoustic backbone and task-specific binary classification heads, facilitating task-conditioned inference across a spectrum of voice-related disorders. This multitask learning (MTL) structure promotes beneficial feature sharing across related disease tasks, potentially enhancing generalization, especially for patients presenting with co-existing conditions. We train and rigorously evaluate our framework on the Bridge2AI-Voice v2.0 dataset~\citep{bensoussan2025bridge2ai,goldberger2000physiobank}, a large-scale, ethically sourced repository of speech samples linked to diagnostic labels for neurological, respiratory, and vocal disorders. Notably, this dataset provides pre-extracted acoustic features and de-identified metadata, aligning perfectly with our privacy-conscious modeling approach that obviates the need for raw audio access. In addition, voice-based analysis lends itself to a wide range of real-world clinical applications. It can be integrated into remote monitoring systems to track patients with chronic or progressive conditions such as heart failure or neurodegenerative diseases. In triage settings, brief voice recordings may assist in early screening and prioritization of at-risk individuals. Moreover, such non-invasive and low-cost tools are particularly valuable in low-resource environments, where access to specialized clinicians or diagnostic infrastructure is limited. Together, these use cases underscore the practical value of speech as a scalable modality for population-level health assessment.

To complement our \SysName~ framework, we further investigate the clinical relevance of established acoustic biomarkers by conducting an auxiliary analysis examining the correlation between the learned deep embeddings from \SysName~ and traditional handcrafted acoustic features. Previous studies show that handcrafted features alone can offer reasonable discrimination across certain disorder subtypes, motivating their use as a reference point for model interpretation \citep{low2020automated,vasquez2019convolutional}. This post hoc analysis reveals measurable similarities between the two, providing additional interpretability into what our model has implicitly learned and showing that it captures clinically relevant vocal patterns in a richer and more nuanced manner than can be described by predefined acoustic features. Moreover, the better performance of \SysName~ compared to models trained directly on handcrafted features suggests that our model leverages these acoustic cues while also learning more complex and task-specific representations.

We summarize our primary contributions as follows:
\begin{itemize}
    \item \textbf{Multi-task and Dual-branch Learning for Speech-Based Disease Classification:} We propose a multitask, dual-branch deep learning framework \SysName~ that operates solely on derived speech representations (MFCCs and spectrograms). This framework supports task-conditioned inference across multiple neurological, respiratory, and vocal disorders, enhancing generalization while adhering to privacy-by-design principles.
    \item \textbf{Robust Benchmarking on the Bridge2AI-Voice Dataset:} We establish a strong and reproducible benchmark on the Bridge2AI-Voice v2.0 dataset using only speech-derived features, without accessing raw audio. Our approach demonstrates significant performance across key clinical tasks (e.g., achieving an F1-score of 0.85 and AUROC of 0.94 for mild cognitive impairment), offering a valuable baseline for future research in privacy-conscious voice-based health assessment.
    \item \textbf{Interpretability through Acoustic Feature Correlation:} Building on the observation that handcrafted acoustic features alone can provide reasonable discrimination for several disorder subtypes, we conduct a post hoc analysis examining the correlation between learned deep embeddings from \SysName~ and these clinically grounded descriptors. This analysis reveals task-specific similarities between our model’s internal representations and established acoustic patterns, enhancing interpretability while also demonstrating that \SysName~ captures richer and more nuanced information beyond what handcrafted features explicitly encode.
\end{itemize}

\section{Related Work}

\subsection{Acoustic Approaches to Vocal Health Assessment}
The utility of voice as a non-invasive modality for diagnosing a wide array of conditions, spanning neurological, respiratory, and voice-related disorders, has garnered significant research interest in recent years \citep{kim2023non,gomez2013characterizing,tong2022respiratory}. Voice-based biomarkers are particularly compelling due to their inherent accessibility, low cost, and sensitivity to subtle alterations in motor and respiratory function. For instance, Parkinson’s disease (PD) frequently manifests through characteristic changes such as monotone prosody and imprecise articulation\citep{brin1992movement,canter1963speech}. Similarly, chronic respiratory diseases like Chronic Obstructive Pulmonary Disease (COPD) are often associated with distinct alterations in respiratory acoustics, potentially indicating deficits in breath control and upper airway coordination \citep{gross2009coordination}. Given the substantial global prevalence of such conditions—COPD alone affects over 300 million people globally~\citep{vogelmeier2017global}, and neurological disorders like PD impact a rapidly growing aging population~\citep{aarsland2021parkinson}—there is an urgent and unmet need for scalable, remote, and non-invasive screening and monitoring tools. Consequently, voice analysis presents a timely and clinically relevant avenue to support early detection and ongoing management across diverse disease domains.

Initial explorations in this field predominantly relied on handcrafted acoustic features—such as jitter, shimmer, harmonic-to-noise ratio (HNR), and formants—coupled with traditional machine learning classifiers like Support Vector Machines (SVMs) and Gaussian Mixture Models (GMMs)~\citep{anagnostopoulos2015features, verde2018voice}. While these seminal studies established foundational links between vocal anomalies and clinical conditions, they often faced challenges in terms of robustness and generalizability across diverse datasets and recording conditions. The advent of deep learning has revolutionized the field, enabling automatic feature extraction directly from acoustic representations like MFCCs and spectrograms. Convolutional Neural Networks (CNNs) and recurrent architectures, such as Bidirectional Long Short-Term Memory networks (BiLSTMs), have proven effective in modeling the complex spatial and temporal patterns inherent in these representations~\citep{Peng2023, alam2025revamping}. More recently, large-scale self-supervised learning (SSL) models, pre-trained on vast amounts of unlabeled audio data, such as wav2vec 2.0 \citep{baevski2020wav2vec} and Whisper~\citep{radford2023robust}, have demonstrated remarkable performance in various speech processing tasks, including applications in medical audio analysis~\citep{Cai2024,violeta2022investigating}.

To leverage complementary information from different acoustic views or modalities, fusion-based architectures have also been explored. For example, multibranch CNN or Temporal Convolutional Network (TCN) models have been applied to tasks like pathological breath sound classification~\citep{zhao2022automatic,kim2024real}. However, many prior works employing fusion tend to use homogeneous encoder structures (e.g., identical CNN backbones for different inputs), potentially limiting their capacity to leverage architecture-specific inductive biases suited for distinct feature types. Furthermore, a common limitation in speech-based disease detection is the assumption of single-label classification, which overlooks the clinical reality that patients often exhibit overlapping symptoms or co-morbidities across multiple systems (e.g., concurrent vocal and respiratory issues).

Our work differentiates itself by proposing a \textit{structurally heterogeneous} dual-branch fusion model. This model processes MFCCs and log-Mel spectrograms through distinct, specialized encoders—ResNet and EfficientNet, respectively. This architectural choice is informed by recent findings demonstrating the unique strengths of these backbones for specific clinical audio tasks: Yu et al.~\cite{yin2024sarnet} highlighted ResNet's efficacy in capturing temporal patterns for dysarthria severity modeling via its residual connections, while Kumar et al.~\cite{kumar2024depression} showcased EfficientNet's advantages in spatial modeling of log-Mel spectrograms for depression detection through compound scaling. By integrating these complementary encoder types within a unified framework, we aim to capture a richer, more comprehensive acoustic representation. Crucially, our model operates exclusively on these derived acoustic features, ensuring privacy-conscious deployment without reliance on raw audio, a critical consideration for clinical applications.

\subsection{Multitask Learning in Clinical AI and Voice Diagnostics}
Multitask learning (MTL) has emerged as a powerful paradigm for simultaneously modeling multiple clinically relevant targets, particularly when these tasks are correlated and can benefit from shared underlying feature representations~\citep{caruana1997multitask}. In the broader healthcare domain, MTL has been successfully applied to diverse problems, including prediction tasks using electronic health records (EHRs), patient outcome modeling, and multimodal clinical reasoning~\citep{niu2025medical, ding2019effectiveness}. Such models typically employ shared encoder layers (backbone) followed by task-specific output heads, often leading to improvements in sample efficiency, model robustness, and generalization performance due to the regularizing effect of shared knowledge.

In the context of voice-based diagnostics, however, the majority of existing models remain confined to single-task setups~\citep{Orozco2016PD,govindu2023early}, often assuming that each speech recording corresponds to a singular disease label. This simplification does not adequately reflect real-world clinical scenarios where patients frequently present with symptoms spanning multiple conditions, such as co-occurring vocal and respiratory disorders. Single-task models may struggle to disentangle these complex, overlapping patterns. MTL offers a natural and effective solution by explicitly modeling shared vocal biomarkers across different tasks while simultaneously learning to retain task-specific nuances necessary for accurate discrimination.

While recent studies have begun to explore the application of MTL to pathological speech analysis, for instance, in jointly assessing voice quality and speech intelligibility \citep{joshy2023dysarthria}, several gaps remain. Many of these approaches still rely on homogeneous input types (e.g., using only spectrograms or only MFCCs for all tasks) or restrict their modeling scope to a single family of disorders (e.g., only neurological conditions). Furthermore, the full potential of leveraging complementary information from diverse acoustic representations within an MTL framework has not been thoroughly explored.

Our research extends this line of inquiry in two significant directions. Firstly, we adopt a dual-branch fusion architecture where MFCCs and spectrograms are processed independently by specialized backbone encoders before fusion. This design allows the model to harness complementary temporal and spectral information more effectively than homogeneous approaches. Secondly, we broaden the diagnostic scope to concurrently address a diverse set of respiratory, vocal, and neurological disorders within a single, unified MTL framework. A critical distinction of our work is its exclusive reliance on speech-derived acoustic features. Unlike many clinical AI models that integrate multimodal inputs such as EHR data or medical imaging, our approach is tailored for scalable, privacy-conscious diagnostics using readily accessible voice data, making it particularly well-suited for remote and resource-constrained settings.

\section{Methodology}
\label{sec:methodology}
\subsection{Problem Definition}
Clinical speech recordings encapsulate a wealth of acoustic information that can serve as indicators for a diverse range of health conditions, including those affecting respiratory, cognitive, and vocal fold function. These recordings are typically elicited via structured vocalization tasks, such as sustained phonation (e.g., holding an /a/ sound) or the reading of standardized passages (e.g., the ``Rainbow Passag''), which are designed to accentuate disorder-specific speech patterns and vocal anomalies.

We frame the diagnostic challenge as a \textit{multi-task binary classification} problem. For any given speech sample, represented by its derived acoustic features, and a specific diagnostic task (e.g., detecting Airway Stenosis), the objective of our model is to predict the presence or absence of the corresponding clinical condition. While each diagnostic task possesses its own binary label space ($y \in \{0, 1\}$), all tasks share the same input modalities derived from the speech signal.

To effectively address the inherent heterogeneity of acoustic cues across different disorders, we formulate our approach as \textit{task-conditional dual-modal learning}. Formally, for each input sample $i$, we have:
\begin{itemize}
    \item $\mathbf{x}_i^{\text{mfcc}} \in \mathbb{R}^{D_M \times T_i}$: The Mel-frequency cepstral coefficient (MFCC) representation, where $D_M$ is the number of MFCC coefficients and $T_i$ is the number of time frames for sample $i$.
    \item $\mathbf{x}_i^{\text{spec}} \in \mathbb{R}^{D_S \times T_i}$: The log-Mel spectrogram representation, where $D_S$ is the number of Mel filter banks and $T_i$ is the number of time frames.
    \item $k \in \{1, \dots, K\}$: The task index, where $K$ is the total number of distinct diagnostic conditions being modeled.
    \item $y_{i,k} \in \{0, 1\}$: The binary ground-truth label for sample $i$ corresponding to task $k$.
\end{itemize}
The goal is to learn a function $f$:
\begin{equation}
f: (\mathbf{x}^{\text{mfcc}}, \mathbf{x}^{\text{spec}}, k) \mapsto \hat{y}_k
\label{eq:problem_def}
\end{equation}
where $\hat{y}_k$ is the predicted probability of the presence of condition $k$. This formulation facilitates parameter sharing across related tasks through a common backbone while maintaining task-specific output layers, thereby enabling nuanced, task-conditioned predictions.

\subsection{Dataset and Input Representations}
\label{sec:dataset}
Our study utilizes a curated subset of the Bridge2AI-Voice v2.0 dataset~\citep{bensoussan2025bridge2ai,goldberger2000physiobank}, a large-scale, ethically sourced voice health database. This dataset encompasses over 19,000 voice recordings from 442 unique participants, collected across five clinical institutions in North America. Each participant performed a standardized set of prompted speech tasks, and these recordings are linked with clinical diagnoses. Critically, to safeguard participant privacy, the publicly accessible version of the dataset exclusively contains derived acoustic representations—including log-Mel spectrograms, MFCCs, and a variety of statistical voice descriptors—with all raw audio data excluded.

For our purpose, we defined nine distinct binary classification tasks, categorized into three major disorder groups:
\begin{enumerate}
    \item \textbf{Respiratory disorders:} Airway Stenosis, Asthma, Chronic Obstructive Pulmonary Disease (COPD).
    \item \textbf{Voice disorders:} Laryngeal Cancer, Vocal Cord Lesions (benign), Spasmodic Dysphonia/Laryngeal Tremor, Vocal Fold Paralysis.
    \item \textbf{Neurological disorders:} Parkinson’s Disease (PD), Alzheimer’s Disease/Mild Cognitive Impairment (AD/MCI).
\end{enumerate}
These specific tasks were selected based on several criteria: sufficient data availability within the Bridge2AI-Voice dataset, strong clinical grounding in prior voice-based research, and the potential for impactful clinical application.

For each binary classification task, positive samples were derived from participants clinically diagnosed with the target condition. 
Negative samples were carefully selected from participants diagnosed with disorders distinct from the target condition, ensuring diagnostic independence between tasks and mitigating the risk of label leakage. 
Each diagnostic condition was thus modeled as an independent binary classification task, with its own definition of positive and negative samples. 
When patients presented with multiple comorbid conditions, each diagnosis was treated as a separate binary decision, and tasks were not assumed to be mutually exclusive. 
As a result, a single voice recording could contribute to multiple task-specific datasets—serving as a positive sample for one condition and a negative or positive sample for others—without introducing methodological conflict. 
This formulation reflected the clinical reality of overlapping diagnoses and supported the multitask architecture’s ability to learn both shared and condition-specific acoustic patterns.

Detailed statistics regarding the distribution of participants and samples across each diagnostic category for training and testing splits are presented in Table~\ref{tab:data_stats_table}. In total, our analytical dataset comprised 11,667 positively labeled recordings originating from 296 unique patients, covering over 49 distinct speech tasks performed by these individuals.

\begin{table*}[htbp]
\caption{Participant and sample statistics for each binary classification task. P = positive class, N = negative class.}
\label{tab:data_stats_table}
\begin{tabular}{@{}llcccccc@{}}
\toprule
\multirow{2}*{\textbf{Disorder}} & \multirow{2}*{\textbf{Subtype Diagnosis}} & \multicolumn{2}{c}{\textbf{Train Participants (\#S)}} & \multicolumn{2}{c}{\textbf{Test Participants (\#S)}} & \multicolumn{2}{c}{\textbf{Total}} \\ \cmidrule(l){3-8} 
 &  & P & N & P & N & P & N \\ \midrule
Respiratory & Airway Stenosis & 42 (1239) & 42 (1527) & 9 (284) & 9 (331) & 51 & 51  \\
 & COPD & 7 (251) & 7 (241) & 4 (130) & 4 (133) & 11 & 11 \\
 & Asthma & 27 (905) & 27 (870) & 6 (214) & 6 (194) & 33 & 33 \\
 & Vocal Fold Paralysis & 25 (875) & 25 (891) & 5 (166) & 5 (155) & 30 & 30 \\ \midrule
Voice & Laryngeal Cancer & 6 (203) & 6 (210) & 2 (70) & 2 (66) & 8 & 8 \\
 & Benign Lesions of the Vocal Cord & 19 (679) & 19 (663) & 5 (174) & 5 (221) & 24 & 24 \\
 & Spasmodic Dysphonia/Laryngeal Tremor & 31 (1091) & 31 (1056) & 7 (235) & 7 (321) & 38 & 38 \\ \midrule
Neurological & Parkinson's & 51 (2284) & 51 (1779) & 10 (874) & 10 (575) & 61 & 61 \\
 & Alzheimer's Disease (AD) and MCI & 34 (1513) & 34 (1094) & 6 (480) & 6 (306) & 40 & 40 \\
\bottomrule
\end{tabular}
\end{table*}

Each voice recording is processed into two primary time-frequency representations, which serve as inputs to our model: \textit{Log-Mel Spectrograms ($\mathbf{x}^{\text{spec}}$):} These capture the spectral energy distribution of the speech signal over time. The resulting power spectrum is then warped onto the Mel scale and log-transformed. For our model, these are shaped as $\mathbb{R}^{1 \times T \times F}$, where $F$ is the number of Mel filters (e.g., 128) and $T$ is the number of time frames. \textit{Mel-Frequency Cepstral Coefficients (MFCCs, $\mathbf{x}^{\text{mfcc}}$):} MFCCs provide a compact representation of the spectral envelope, effectively capturing smooth vocal tract dynamics. They are derived by applying a Discrete Cosine Transform (DCT) to the log-Mel spectrum. We use 60 MFCCs (including C0), shaped as $\mathbb{R}^{1 \times T \times 60}$.

\begin{figure*}[htbp]
  \centering
  \includegraphics[width=\textwidth]{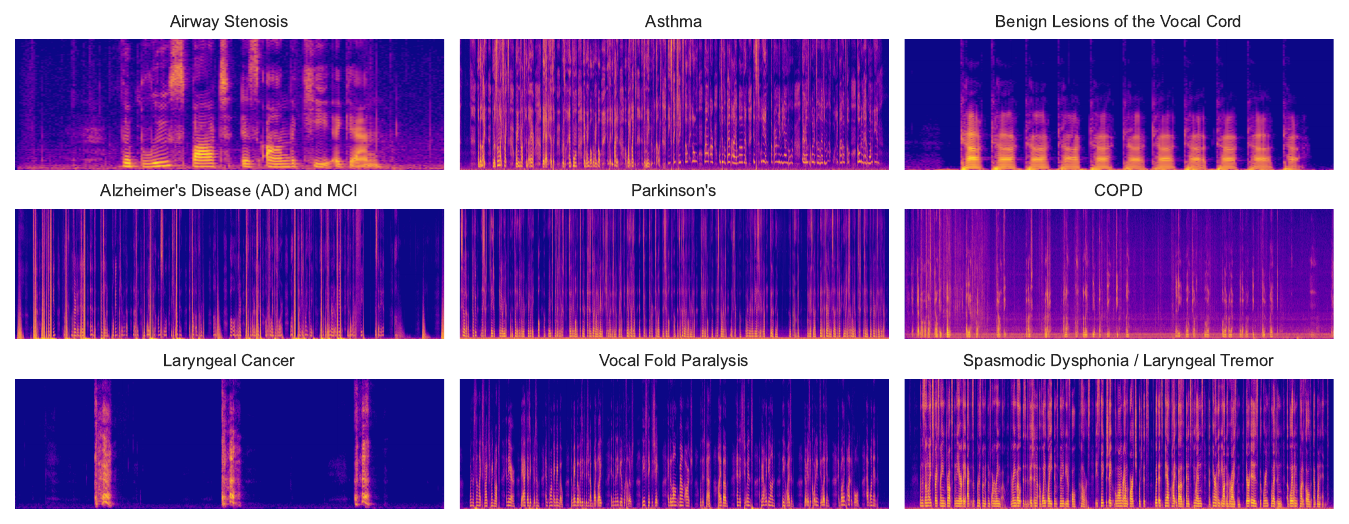}
  \caption{Representative log-Mel spectrograms across nine voice-related disorders. Each panel illustrates a speech recording from one participant, showcasing condition-specific spectral patterns such as disrupted harmonics, reduced energy in higher frequencies, or atypical temporal structure. These visual differences underscore the diagnostic relevance of time–frequency representations in voice-based disease detection.}
  \label{fig:input_representations}
\end{figure*}

To offer a qualitative insight into the acoustic diversity across different conditions, Figure~\ref{fig:input_representations} visualizes log-Mel spectrograms for the nine targeted voice-related disorders. Each panel corresponds to a speech sample from a single participant performing a specific task, highlighting how disease-specific vocal characteristics (e.g., disrupted harmonics, altered energy distribution, atypical temporal structure) manifest in the time-frequency domain.

To ensure robust generalization assessment and prevent data leakage, we implement \textit{patient-level stratified splits}. This means all recordings from a single patient are assigned exclusively to either the training set or the test set for each task, stratified by the diagnostic label. Further details on the experimental setup are provided in Section~\ref{sec:experiments}.

\subsection{Multitask Model Architecture}
\label{sec:model_architecture} 

To effectively model the complex acoustic manifestations associated with diverse vocal disorders, we design a \textbf{multi-task dual-branch neural network}. This architecture is designed to integrate information from both log-Mel spectrograms and MFCCs.

\subsubsection{Modality-Specific Encoders}
Input audio segments, preprocessed into the key representations described in Section~\ref{sec:dataset}, are fed into separate, specialized encoder backbones:

\begin{itemize}
    \item \textbf{Spectrogram Encoder ($f_{\text{spec}}$):} An EfficientNet-B0~\cite{tan2019efficientnet} architecture is employed to process the log-Mel spectrograms ($\mathbf{X}^{\text{spec}}$). To align with the standard 3-channel input expected by the ImageNet-pretrained model, the single-channel spectrogram is passed through an initial 2D convolutional layer that projects it to 3 channels before entering the EfficientNet backbone.
    \item \textbf{MFCC Encoder ($f_{\text{mfcc}}$):} A ResNet18~\cite{he2016deep} architecture is used for the MFCC inputs ($\mathbf{X}^{\text{mfcc}}$). The first convolutional layer of the ResNet18 is modified to accept single-channel input, corresponding to the MFCC feature dimension over time.
\end{itemize}

The choice of these distinct encoders allows leveraging the specific strengths of EfficientNet for rich 2D spatial feature extraction from spectrograms and ResNet for capturing temporal patterns from MFCC sequences. Each encoder $f_{\text{modality}}$ outputs a modality-specific embedding:

\begin{equation}
\mathbf{h}_{\text{mfcc}} = f_{\text{mfcc}}(\mathbf{X}^{\text{mfcc}}) \in \mathbb{R}^{D_{\text{emb,mfcc}}}, \quad \mathbf{h}_{\text{spec}} = f_{\text{spec}}(\mathbf{X}^{\text{spec}}) \in \mathbb{R}^{D_{\text{emb,spec}}}
\end{equation}

\noindent where $D_{\text{emb,mfcc}}=512$ (output dimension of ResNet18 average pooling layer) and $D_{\text{emb,spec}}=1280$ (output dimension of EfficientNet-B0 average pooling layer).

\subsubsection{Feature Fusion and Shared Representation Layer}
The modality-specific embeddings are subsequently concatenated to form a joint feature representation:

\begin{equation}
\mathbf{h}_{\text{fused}} = \text{Concat}(\mathbf{h}_{\text{mfcc}}, \mathbf{h}_{\text{spec}}) \in \mathbb{R}^{(D_{\text{emb,mfcc}} + D_{\text{emb,spec}})}
\end{equation}

In our case, $\mathbf{h}_{\text{fused}} \in \mathbb{R}^{1792}$. This fused representation is then processed by a shared transformation module, $g(\cdot)$, which consists of a linear projection layer (reducing dimensionality from 1792 to 512), followed by batch normalization, a LeakyReLU activation function ($\alpha=0.1$), and dropout. This module produces a task-agnostic shared latent vector $\mathbf{z}$:

\begin{equation}
\mathbf{z} = g(\mathbf{h}_{\text{fused}}) \in \mathbb{R}^{512}
\end{equation}

This shared representation layer aims to learn common underlying acoustic patterns relevant across multiple disorders.

\subsubsection{Task-Specific Prediction Heads}
For each of the $K$ diagnostic tasks (here $K=9$), a dedicated binary classification head, $h^{(k)}(\cdot)$, is appended to the shared representation layer. Each head is implemented as a two-layer Multi-Layer Perceptron (MLP):
\begin{itemize}
    \item An initial linear layer transforming $\mathbf{z}$ from 512 to 128 dimensions, followed by batch normalization and LeakyReLU activation ($\alpha=0.1$).
    \item A final linear layer projecting the 128-dimensional representation to a single logit for binary classification.
    \item Dropout regularization is applied within each head.
\end{itemize}
Each head $h^{(k)}$ operates on the shared representation $\mathbf{z}$ to produce a task-specific prediction $\hat{y}^{(k)}$:
\begin{equation}
\hat{y}^{(k)} = \sigma(\text{MLP}^{(k)}(\mathbf{z})) = \sigma(\mathbf{W}^{(k)}\mathbf{z}_{\text{head\_in}}^{(k)} + \mathbf{b}^{(k)})
\end{equation}

\noindent where $\sigma(\cdot)$ is the sigmoid activation function, converting the logit to a probability. To address potential class imbalance inherent in clinical datasets, we employ a weighted binary cross-entropy (BCE) loss for each task:

\begin{equation}
\mathcal{L}^{(k)} = - \left( w_1^{(k)} y^{(k)} \log \hat{y}^{(k)} + w_0^{(k)} (1 - y^{(k)}) \log (1 - \hat{y}^{(k)}) \right)
\end{equation}

\noindent where $w_1^{(k)}$ and $w_0^{(k)}$ are weights assigned to the positive and negative classes for task $k$, respectively, typically set inversely proportional to class frequencies. The overall training objective is the sum of losses across all $K$ tasks:
\begin{equation}
\mathcal{L}_{\text{total}} = \sum_{k=1}^{K} \mathcal{L}^{(k)}
\end{equation}

The training strategy and related details are provided in Section~\ref{sec:experimental_setup}.

\section{Experiments}
\label{sec:experiments}

\subsection{Experimental Setup}
\label{sec:experimental_setup} 

This section details the specific configurations and choices made for training and evaluating our proposed models. The dataset, task definitions, and core model architecture have been described in Section~\ref{sec:methodology}.

\textbf{Baselines and Model Configurations:}
To comprehensively assess the efficacy of our proposed multitask dual-modal framework (\SysName), we compare it against several strong baselines:
\begin{itemize}
    \item \textbf{Single-Modality Deep Learning Models:}
    \begin{itemize}
        \item \textbf{$\text{EN}_s$ (EfficientNet-Spectrogram)}: An EfficientNet-B0 model trained solely on log-Mel spectrogram inputs.
        \item \textbf{$\text{EN}_m$ (EfficientNet-MFCC)}: An EfficientNet-B0 model trained solely on MFCC inputs.
        \item \textbf{$\text{RN}_s$ (ResNet-Spectrogram)}: A ResNet18 model trained solely on log-Mel spectrogram inputs.
        \item \textbf{$\text{RN}_m$ (ResNet-MFCC)}: A ResNet18 model trained solely on MFCC inputs.
    \end{itemize}
    These single-modality models share the same encoder architecture as their respective branches in the MTL model but are trained in a single-task fashion for each of the nine disorders independently. Their outputs are then averaged or aggregated for comparison at the category level.
    \item \textbf{MLP with Handcrafted Features (\textbf{MLP})}: A shallow Multi-Layer Perceptron model trained on a comprehensive set of 131 traditional, handcrafted acoustic features (e.g., jitter, shimmer, HNR, formants). This baseline serves to evaluate the performance achievable with established acoustic biomarkers and provides context for our deep learning approaches.
\end{itemize}
Our proposed model, as detailed in Section~\ref{sec:model_architecture}, combines ResNet18 for MFCCs and EfficientNet-B0 for spectrograms, learning a shared representation across all nine diagnostic tasks simultaneously.

\textbf{Training Details:}
All deep learning models (MTL and single-modality baselines) were trained using the AdamW optimizer with an initial learning rate of $1 \times 10^{-4}$ and a weight decay of $1 \times 10^{-5}$. A cosine annealing learning rate schedule was employed. We used a batch size of 108. For the MTL model, we implemented a \textit{multi-task balanced batch sampler}, ensuring each mini-batch contained 6 positive and 6 negative samples per task (totaling $9 \text{ tasks} \times 12 \text{ samples/task} = 108$ samples per batch). Sampling with replacement was used when insufficient unique samples were available for a given task within an epoch. All models were trained for a maximum of 40 epochs, with dropout regularization (rate of 0.3) applied to fully connected layers. Gradient clipping (norm 1.0) was used to ensure stability. The MLP baseline was trained using similar optimization hyperparameters.

\textbf{Data Augmentation:}
To enhance model generalization, distinct data augmentation techniques were applied during training. For MFCC inputs, we appliep zero-mean Gaussian noise $\mathcal{N}(0, \sigma^2)$, where $\sigma = 0.01$. Further, we employed frequency masking, where 8 consecutive frequency bins were masked, and time masking, where 20 consecutive time frames were masked. For log-Mel spectrogram inputs, we utilized SpecAugment~\citep{park2019specaugment}, which involved frequency masking (masking 15\% of the frequency bins) and time masking (masking 15\% of the time frames), in addition to the time warping procedure detailed in Section~\ref{sec:model_architecture}. These augmentations were applied stochastically exclusively during training.

\textbf{Evaluation Metrics:}
The primary evaluation metric reported for all models and tasks is the Area Under the Receiver Operating Characteristic Curve (AUROC). AUROC provides a robust measure of classification performance, particularly well-suited for medical diagnostic tasks that may exhibit class imbalance. To ensure robustness and statistical significance, all experiments were conducted with five independent runs.

\subsection{Results}
\label{sec:results}
In this section, we present a comprehensive evaluation of our proposed \SysName~ framework against various baselines across the nine defined diagnostic tasks. For completeness, a SHAP-based analysis of handcrafted features is included in Appendix~\ref{sec:appendix_shap}.

\subsubsection{Comparison with Single-Task Single-Modality and MLP Baselines}
\label{sec:comparison_baselines}

Table~\ref{tab:highlevel-auroc} summarizes the AUROC performance (mean $\pm$ standard deviation) of our \SysName~ model compared to single-task, single-modality deep learning baselines ($\text{EN}_m$, $\text{EN}_s$, $\text{RN}_m$, $\text{RN}_s$) and the MLP model trained on handcrafted features. Performance is aggregated at the level of three high-level disorder categories: Respiratory, Voice, and Neurological, as well as an overall average across all tasks. The consistently low standard deviations (< 0.20 for most metrics) indicate stable training performance across different random initializations.

Each baseline model, including the MLP trained on handcrafted features, was trained independently for each task using a single type of acoustic representation—either MFCCs, spectrograms, or handcrafted features—without any task-sharing or modality fusion.

Our model consistently outperforms all single-task baselines across all aggregated disorder types, achieving the highest overall AUROC of $0.78 \pm 0.14$. The model demonstrates a particularly strong discriminative capacity for neurological disorders ($0.89 \pm 0.11$), followed by voice disorders ($0.76 \pm 0.08$) and respiratory disorders ($0.74 \pm 0.18$).

The advantages of the \SysName~ are most pronounced for voice disorders, where it achieves an AUROC of $0.76 \pm 0.08$, a substantial improvement of $0.05$ AUROC points over the best-performing signle-task single-modal baseline in this category ($\text{EN}_s$ at $0.71 \pm 0.09$). For neurological disorders, our model's AUROC of $0.89 \pm 0.11$ surpasses all alternatives, including the competitive MLP baseline ($0.80 \pm 0.09$). While improvements for respiratory disorders are more modest, \SysName~ ($0.74 \pm 0.18$) still offers consistent gains over single-modal approaches, which cluster around AUROCs of $0.70-0.73$.

These results underscore the effectiveness of our multi-task learning strategy. By leveraging shared representations across related disorders and integrating complementary information through dual-modal fusion (MFCCs and spectrograms), our model achieves superior generalization and discriminative power, particularly for conditions intrinsically linked to voice production mechanisms.

\begin{table*}[t]
\centering
\caption{AUROC (mean $\pm$ std over 5 runs) for each disorder category. Baseline models use single-task learning with individual modalities (MFCC, spectrogram, or handcrafted features).}

\label{tab:highlevel-auroc}
\begin{tabular}{lcccccc}
\toprule
\textbf{Disorder Type} & \textbf{MLP} & \textbf{$\text{EN}_m$} & \textbf{$\text{EN}_s$} & \textbf{$\text{RN}_m$} & \textbf{$\text{RN}_s$} & \textbf{\SysName} \\
\midrule
Respiratory  & 0.71 ± 0.15 & 0.73 ± 0.15 & 0.70 ± 0.12 & 0.73 ± 0.14 & 0.72 ± 0.16 & \textbf{0.74 ± 0.18} \\
Voice        & 0.67 ± 0.09 & 0.66 ± 0.11 & 0.71 ± 0.09 & 0.63 ± 0.12 & 0.67 ± 0.07 & \textbf{0.76 ± 0.08} \\
Neurological & 0.80 ± 0.09 & 0.76 ± 0.01 & 0.77 ± 0.08 & 0.74 ± 0.06 & 0.73 ± 0.04 & \textbf{0.89 ± 0.11} \\
\midrule
All          & 0.71 ± 0.10 & 0.70 ± 0.11 & 0.71 ± 0.09 & 0.67 ± 0.11 & 0.69 ± 0.09 & \textbf{0.78 ± 0.14} \\
\bottomrule
\end{tabular}
\end{table*}

\begin{figure*}[htbp]
  \centering
\includegraphics[width=.9\textwidth]{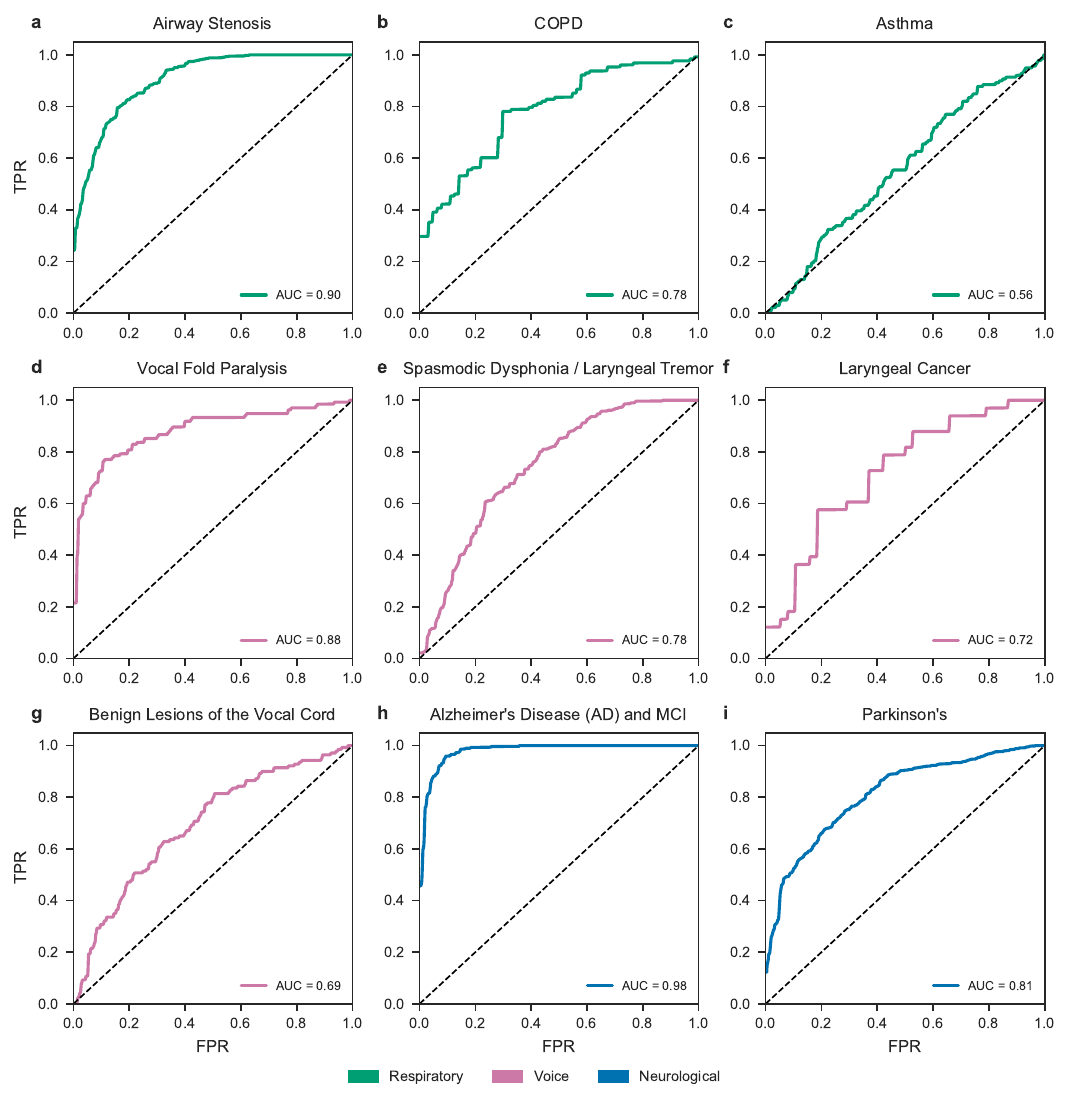}
  \caption{Task-level ROC curves of the proposed multitask voice-based disease classifier across nine subtype disorders.}
  \label{fig:roc_curves}
\end{figure*}

\subsubsection{Task-Specific Performance Analysis}
\label{sec:task_specific_roc}
Figure~\ref{fig:roc_curves} provides a detailed visualization of the \SysName~ task-level performance through Receiver Operating Characteristic (ROC) curves for each of the nine subtype disorders.Results are shown for the model with the best overall multi-task performance.These curves reveal distinct performance patterns across and within disease categories, offering insights into the model's discriminative capabilities for specific conditions.
\begin{itemize}
    \item \textbf{Neurological Disorders (Blue ROC Curves):} Exhibit exceptional performance overall. Alzheimer's Disease/MCI detection achieves the highest discrimination (AUROC = $0.98$), indicating very strong separability. Parkinson's Disease classification also demonstrates robust capability (AUROC = $0.81$).
    \item \textbf{Respiratory Disorders (Green ROC Curves):} Show more variable performance. Airway Stenosis classification yields excellent results (AUROC = $0.90$), while COPD detection is also strong (AUROC = $0.78$). Asthma, however, presents the most challenging task (AUROC = $0.56$), likely due to the subtlety or variability of its acoustic manifestations in the speech data.
    \item \textbf{Voice Disorders (Purple ROC Curves):} Generally demonstrate intermediate to strong performance. Vocal Fold Paralysis classification is a standout in this category (AUROC = $0.88$), indicating clear acoustic markers are learned by the model. Other voice disorders show moderate to good discrimination.
\end{itemize}

\begin{table*}[t]
\centering
\caption{AUROC comparison between single-task self-supervised baselines and the proposed multi-task learning (MTL) model across subtype disorders (mean ± std over 5 runs).}
\label{tab:sup_ssl_comparison_table}
\begin{tabular}{@{}lccc|c@{}}
\toprule
\textbf{Subtype Task} & \textbf{Whisper} \cite{radford2023robust} & \textbf{Wav2vec-BERT} \cite{barrault2023seamless} & \textbf{M-CTC-T} \cite{lugosch2022pseudo} & \textbf{\SysName} \\
\midrule
Airway Stenosis & 0.84 ± 0.02 & 0.80 ± 0.03 & 0.78 ± 0.03 & \textbf{0.89 ± 0.01} \\
Asthma & 0.53 ± 0.03 & \textbf{0.56 ± 0.03} & 0.53 ± 0.03 & 0.54 ± 0.03 \\
Benign Lesions of the Vocal Cord & 0.51 ± 0.03 & 0.64 ± 0.03 & 0.60 ± 0.03 & \textbf{0.70 ± 0.01} \\
COPD & 0.67 ± 0.04 & 0.74 ± 0.03 & 0.75 ± 0.03 & \textbf{0.79 ± 0.03} \\
Laryngeal Cancer / Pre-cancerous & 0.70 ± 0.03 & 0.69 ± 0.03 & 0.67 ± 0.03 & \textbf{0.71 ± 0.04} \\
Alzheimer's Disease (AD) and MCI & 0.91 ± 0.02 & 0.89 ± 0.01 & 0.88 ± 0.02 & \textbf{0.97 ± 0.01} \\
Spasmodic Dysphonia / Laryngeal Tremor & 0.72 ± 0.02 & 0.73 ± 0.03 & 0.72 ± 0.02 & \textbf{0.77 ± 0.02} \\
Vocal Fold Paralysis & 0.83 ± 0.02 & 0.78 ± 0.03 & 0.76 ± 0.03 & \textbf{0.87 ± 0.01} \\
Parkinson's & 0.79 ± 0.04 & 0.78 ± 0.04 & 0.74 ± 0.05 & \textbf{0.81 ± 0.06} \\
\bottomrule
\end{tabular}
\end{table*}

The variation in ROC curve shapes and AUROC values across tasks provides valuable information for potential clinical deployment strategies. High-performing tasks like AD/MCI and Airway Stenosis detection could support sensitive screening applications. Conversely, more challenging conditions like Asthma may require more conservative decision thresholds or integration with other diagnostic modalities (in particular breathing sounds), reflecting the underlying acoustic pathophysiology and guiding targeted clinical implementation.

\subsubsection{Comparison with Self-Supervised Learning Baselines}
\label{sec:comparison_ssl}
To further contextualize the performance of our supervised \SysName~ framework, we compare it against contemporary self-supervised learning (SSL) models commonly used in speech processing. Table~\ref{tab:sup_ssl_comparison_table} presents a task-by-task AUROC comparison with Whisper-small~\citep{radford2023robust}, Wav2Vec2-BERT~\citep{barrault2023seamless}, and M-CTC-T~\citep{lugosch2022pseudo}. For fair comparison, these SSL models were fine-tuned on our specific downstream tasks using the same training data splits and by their design only utilize spectrograms.

The results indicate that our model outperforms these SSL baselines on the majority of disorder subtypes (7 out of 9 tasks). Notably, \SysName~ achieves exceptional performance on Alzheimer's Disease/MCI detection ($0.97 \pm 0.01$) and Airway Stenosis classification ($0.89 \pm 0.01$), surpassing the SSL models by a considerable margin.

While SSL models benefit from pretraining on large-scale, diverse speech corpora, they may not always capture the fine-grained, domain-specific acoustic biomarkers crucial for accurate medical diagnosis without extensive task-specific fine-tuning. In contrast, our \SysName~ framework, with its task-specific heads, dual-modal fusion tailored to relevant acoustic features, and supervised training on in-domain clinical data, learns more discriminative and clinically pertinent representations for voice-based disease classification. This highlights the current advantage of supervised, specialized architectures for these specific clinical tasks when sufficient labeled data is available.

\subsubsection{Latent Space Visualization}
\label{sec:latent_space_viz}

\begin{figure}[htbp]
\centering
\includegraphics[width=0.8\textwidth]{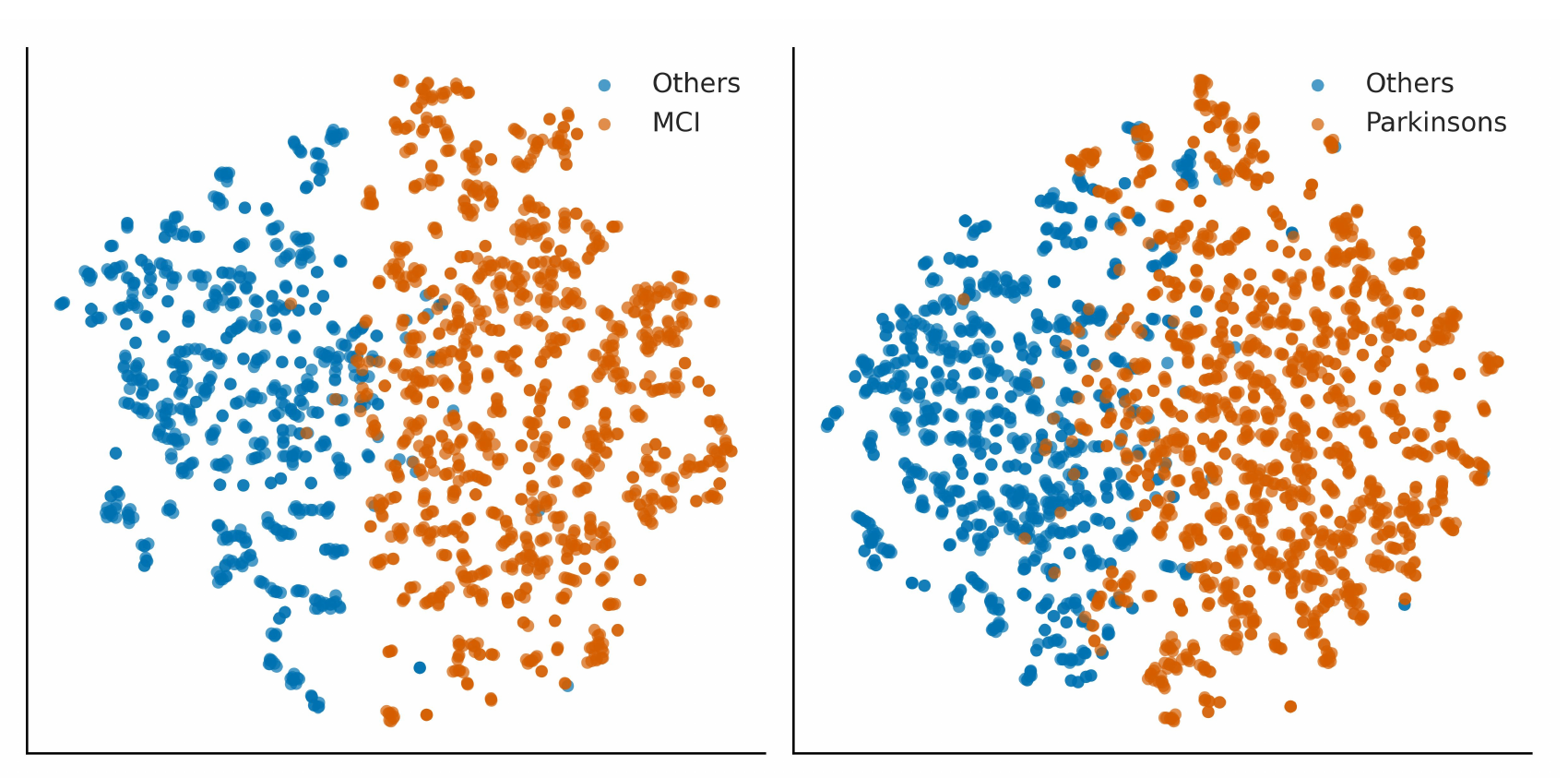}
\caption{t-SNE visualization of penultimate-layer embeddings from the our model. Left: MCI vs. control; Right: Parkinson’s vs. control. Clear separation in the MCI task suggests strong latent discriminability, while Parkinson’s exhibits more subtle structure.}
\label{fig:tsne_plot}
\end{figure}

To qualitatively assess the discriminative capacity of the representations learned by our model, we visualize the penultimate-layer embeddings (the 512-dimensional vector $\mathbf{z}$) using t-distributed Stochastic Neighbor Embedding (t-SNE)~\citep{van2008visualizing}. Figure~\ref{fig:tsne_plot} presents these visualizations for two representative tasks: Alzheimer's Disease/MCI (left panel) and Parkinson's Disease (right panel), comparing embeddings from positive (disease) cases against negative (control) cases.

The t-SNE plot for AD/MCI reveals a clear and compact separation of positive and negative samples, with a distinct boundary between the two groups. This strong separation in the learned latent space aligns with the exceptionally high AUROC ($0.97$) observed for this task, suggesting that the model effectively captures stable and highly discriminative patterns for AD/MCI.

In contrast, the t-SNE visualization for Parkinson's Disease shows a slightly more diffuse cluster boundary. While distinct clusters for positive and negative cases are still evident, there is greater intermingling compared to the AD/MCI plot. This suggests that while the model learns condition-specific signals for Parkinson's Disease, these signals might exhibit greater variability or subtlety, consistent with the more moderate AUROC of $0.81$ for this task.

Overall, these t-SNE visualizations provide compelling visual evidence that our dual-modal architecture successfully encodes task-relevant acoustic features, learning a latent space that exhibits strong separability for clinically relevant conditions, thereby corroborating the quantitative performance metrics.

\subsection{Ablation Study and Component Analysis}
\label{sec:ablation_study}

\begin{table*}[t]
\centering
\caption{AUROC (mean ± std over 5 runs) on each subtype disorder classification task. Baselines use single-task learning while~\SysName~employs multi-task learning across all model-input combinations.}
\label{tab:subtype-auroc-all}
\resizebox{0.92\textwidth}{!}{
\begin{tabular}{lccccc|c}
\toprule
\textbf{Subtype Task} & \textbf{$\text{EN}_m$} & \textbf{$\text{EN}_s$} & \textbf{$\text{RN}_m$} & \textbf{$\text{RN}_s$} & \textbf{MLP} & \textbf{\SysName} \\
\midrule
Airway Stenosis & 0.80 ± 0.03 & 0.80 ± 0.04 & 0.83 ± 0.03 & 0.85 ± 0.04 & 0.84 ± 0.02 & \textbf{0.89 ± 0.01} \\
Alzheimer's Disease (AD) and MCI & 0.77 ± 0.01 & 0.83 ± 0.01 & 0.78 ± 0.01 & 0.75 ± 0.02 & 0.86 ± 0.01 & \textbf{0.97 ± 0.01} \\
Asthma & 0.55 ± 0.03 & \textbf{0.57 ± 0.03} & 0.56 ± 0.04 & 0.54 ± 0.01 & 0.55 ± 0.03 & 0.54 ± 0.03 \\
Benign Lesions of the Vocal Cord & 0.65 ± 0.02 & 0.67 ± 0.01 & 0.61 ± 0.02 & 0.62 ± 0.03 & 0.59 ± 0.05 & \textbf{0.70 ± 0.01} \\
COPD & \textbf{0.83 ± 0.03} & 0.72 ± 0.08 & 0.79 ± 0.05 & 0.76 ± 0.04 & 0.74 ± 0.09 & 0.79 ± 0.03 \\
Laryngeal Cancer / Pre-cancerous & 0.52 ± 0.06 & 0.60 ± 0.11 & 0.46 ± 0.14 & 0.59 ± 0.08 & 0.58 ± 0.12 & \textbf{0.71 ± 0.04} \\
Parkinson's & 0.75 ± 0.02 & 0.71 ± 0.03 & 0.69 ± 0.04 & 0.70 ± 0.03 & 0.73 ± 0.01 & \textbf{0.81 ± 0.06} \\
Spasmodic Dysphonia / Laryngeal Tremor & 0.77 ± 0.01 & 0.80 ± 0.02 & 0.75 ± 0.03 & 0.75 ± 0.04 & 0.77 ± 0.01 & \textbf{0.77 ± 0.02} \\
Vocal Fold Paralysis & 0.66 ± 0.09 & 0.77 ± 0.02 & 0.68 ± 0.04 & 0.70 ± 0.14 & 0.73 ± 0.01 & \textbf{0.87 ± 0.01} \\
\bottomrule
\end{tabular}
}
\end{table*}

To dissect the contributions of our framework's components, we conducted an ablation study comparing our dual-modal multi-task model against single-modality models ($\text{EN}_m$, $\text{EN}_s$, $\text{RN}_m$, $\text{RN}_s$) and an MLP baseline across all nine tasks. Detailed AUROC scores are provided in Table~\ref{tab:subtype-auroc-all}.

\subsubsection{Impact of Multi-Task Learning (MTL)}
\label{sec:ablation_mtl_effectiveness}
\SysName~ framework demonstrates superior performance by outperforming single-task baselines on 7 of 9 tasks. The benefits are most notable for tasks like \textit{Airway Stenosis} (AUROC $0.89 \pm 0.01$ vs. best single-modality $0.85 \pm 0.04$), \textit{Vocal Fold Paralysis} ($0.87 \pm 0.01$ vs. $0.77 \pm 0.02$), and \textit{Alzheimer’s Disease/MCI} ($0.97 \pm 0.01$ vs. $0.86 \pm 0.01$ for MLP). These improvements, also seen in moderately resourced tasks (e.g., Parkinson's, laryngeal cancer), affirm that shared supervision in \SysName fosters robust representation learning, particularly with limited task-specific labeled data.

\begin{figure*}[t]
\centering
\includegraphics[width=\textwidth]{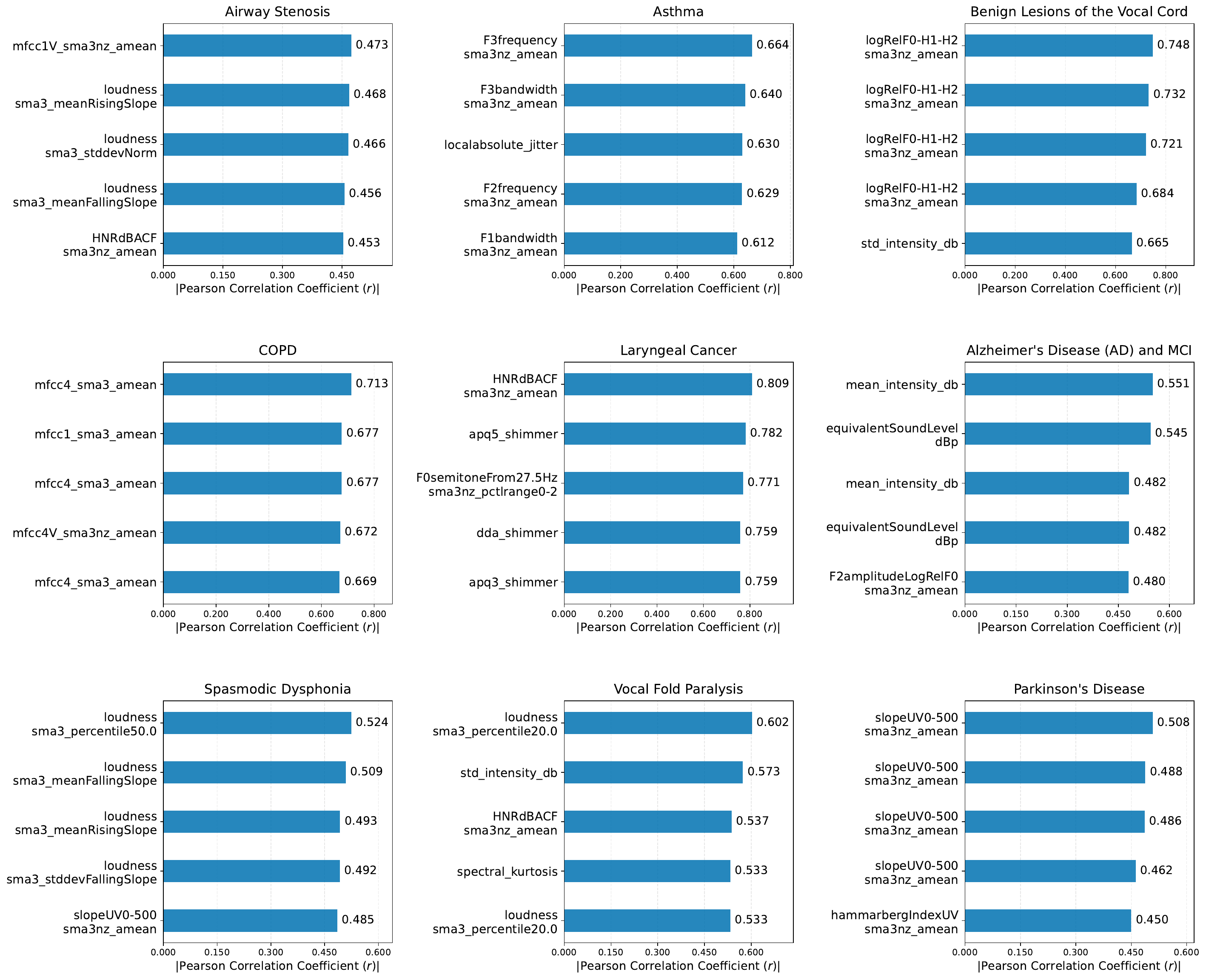}
\caption{Top-5 handcrafted acoustic features most correlated with model embeddings across clinical tasks. For each task, bars show the absolute Pearson correlation between acoustic features and their most strongly associated dimension in the model's final-layer embeddings (computed on test set).
}
\label{fig:corr_bar_all}
\end{figure*}

\subsubsection{Complementarity of Acoustic Modalities (MFCCs vs. Spectrograms)}
\label{sec:ablation_modality}
Analysis of single-modality performance reveals task-specific strengths: MFCC-based models often excel for \textit{neurological disorders} due to sensitivity to temporal and prosodic features, while \textit{respiratory and structural voice disorders} show mixed preferences. Critically, our dual-modal model surpasses the best single-modality configuration on 8 of 9 tasks, with AUROC gains ranging up to $0.19$ points (e.g., Laryngeal Cancer). This underscores that MFCCs (capturing spectral envelope and dynamics) and spectrograms (rich time-frequency details) provide complementary acoustic information, and their fusion yields more discriminative representations than either modality alone, with Spasmodic Dysphonia being a minor exception.

\subsubsection{Influence of Task-Intrinsic Difficulty on Performance}
\label{sec:ablation_task_difficulty}
A consistent performance hierarchy across tasks is observed in all models. \textit{High-performing tasks} like Alzheimer’s Disease/MCI (AUROC $0.97$) and Airway Stenosis ($0.89$) suggest distinct acoustic biomarkers. \textit{Moderately-performing tasks} include Parkinson’s Disease ($0.81$). \textit{Asthma remains a low-performing task} ($0.54$), likely due to subtle or variable acoustic cues, or data ambiguities, rather than limitations of the model itself. As an episodic condition, participants may have been asymptomatic during recording, reducing the presence of characteristic markers. The absence of healthy controls further complicates discrimination by requiring differentiation from other respiratory disorders with overlapping acoustic profiles. These factors suggest limited diagnostic utility of voice for conditions with intermittent symptoms or similar acoustic phenotypes. This indicates that task-intrinsic difficulty is a primary factor constraining performance for certain conditions.

\subsection{Correlation Analysis between Model Embeddings and Acoustic Biomarkers}
\label{sec:correlation_analysis}

To further examine the relationship between learned representations and established acoustic biomarkers within our multi-task framework, we conducted a post hoc correlation analysis between the learned deep embeddings and traditional handcrafted acoustic features. Specifically, we used the 128-dimensional embeddings from the final fused layer of the \SysName~ model on the test set and computed Pearson correlation coefficients with all 131 handcrafted acoustic features extracted from each audio sample. These features, derived using openSMILE and Praat toolkits, include a broad spectrum of prosodic, spectral, and voice quality descriptors such as \texttt{F0\_sma\_stddev} (pitch variability), \texttt{logHNR\_sma3nz\_stddevNorm} (harmonic-to-noise ratio), and \texttt{slopeUV0-500\_sma3nz\_amean} (spectral slope in the low-frequency band).

Figure~\ref{fig:corr_bar_all} presents the top-5 acoustic features most correlated with the learned embeddings for each diagnostic task. These results reveal meaningful task-specific associations between clinically interpretable acoustic attributes and internal model representations, suggesting that \SysName~ implicitly captures patterns consistent with domain knowledge—even without explicit access to engineered features.

In several tasks, we observed strong correlations between learned deep embeddings and clinically meaningful handcrafted acoustic features. In Laryngeal cancer, the highest correlations were found with \texttt{HNRdBACF\_sma3nz\_amean} ($r = 0.809$) and the other feature \texttt{F3frequency\_sma3nz\_amean} ($r = 0.782$), which reflect the harmonic-to-noise ratio and third formant frequency, critical indicators of spectral stability and vocal tract configuration often disrupted in malignant laryngeal conditions. For Benign Lesions in the Vocal Cord task, features such as \texttt{ logRelF0-H1-H2\_sma3nz\_amean} showed high correlation values ($r = 0.748$, $0.732$ and $0.721$ in variants, aligned with irregularities at the glottal source commonly associated with benign structural lesions. In COPD, MFCC-based characteristics, such as \texttt{ mfcc4\_sma3\_amean} ($r = 0.713$), exhibited strong correlations, suggesting that the model captures spectral envelope patterns potentially linked to altered airflow and resonance seen in chronic pulmonary conditions. For Vocal Fold Paralysis, features like \texttt{loudness\_sma3\_percentile20.0} ($r = 0.602$) and \texttt{std\_intensity\_db} ($r = 0.573$) were most correlated, corresponding to reduced vocal intensity and instability—hallmarks of glottal incompetence. Even in Parkinson's disease, where shows moderate correlations, consistent alignment with the features such as \texttt{ slopeUV0-500\_sma3nz\_amean} variants ($r \approx 0.46$–$0.51$) supports the model’s sensitivity to subtle prosodic deficits such as reduced pitch modulation and breathiness, which are typical in Parkinsonian speech.

Rather than quantifying feature importance, this correlation analysis assesses the extent to which the model's learned representations reflect clinically grounded acoustic attributes. The observed task-specific alignments reinforce the clinical plausibility of the internal embeddings.

Importantly, while a lightweight MLP trained on handcrafted acoustic features demonstrates meaningful discriminative performance and captures clinically relevant voice characteristics, our \SysName~ model consistently outperforms it. This performance gap suggests that \SysName~ not only encodes similar diagnostic cues but also learns more comprehensive and nuanced vocal patterns directly from derived acoustic representations—highlighting the value of end-to-end learning beyond manual feature engineering.

We further provide SHAP-based analysis in Appendix~\ref{sec:appendix_shap}, which confirms that many of the highly correlated acoustic features also contribute to disorder detection. Together, these findings suggest that \SysName~ captures clinically meaningful voice patterns while offering richer and more expressive representations than traditional approaches.

\section{Conclusion}
\label{sec:conclusion}
In this work, we introduced \SysName, a multi-task, dual-modal deep learning framework that effectively detects diverse neurological, respiratory, and voice disorders by operating on speech derived MFCC and spectrogram features. Evaluated on the large-scale Bridge2AI-Voice dataset, \SysName~ demonstrated superior performance over single-task, single-modal and self-supervised baselines, with its dual-modal fusion yielding consistent 5--19\% performance gains.
A post hoc analysis revealed that the model’s learned embeddings correlate with clinically meaningful handcrafted acoustic features, offering interpretability into the vocal patterns \SysName~ implicitly captures.
Our findings underscore the significant potential of scalable, non-invasive voice-based diagnostics, establishing a strong foundation for clinical applications. Future work will focus on integrating self-supervised pre-training to leverage unlabeled voice data and incorporating uncertainty quantification to enhance clinical decision-making, ultimately paving the way for deployment-ready AI systems in healthcare practice.

\section*{Acknowledgments}
This work was supported by the AiNed Fellowship Grant awarded to A.S. We acknowledge the use of the Dutch National Supercomputer Snellius for essential computational tasks.

\clearpage

\bibliographystyle{ACM-Reference-Format}
\bibliography{main}

\clearpage

\appendix
\section*{Appendix}
\addcontentsline{toc}{section}{Appendix}

\section{SHAP-Based Feature Attribution for Acoustic Biomarkers}
\label{sec:appendix_shap}

To complement our primary deep learning framework and contextualize the findings from Section~\ref{sec:correlation_analysis}, which revealed correlations between learned embeddings and handcrafted acoustic features, we conducted an auxiliary SHAP-based analysis. This involved training a lightweight Multi-Layer Perceptron (MLP) model on the same set of handcrafted acoustic features used in the correlation study.

The objectives of this analysis were twofold: (1) to independently validate whether traditional, interpretable acoustic features are predictive of voice-related disorders, and (2) to identify which specific acoustic descriptors are most influential for each classification task using SHAP (SHapley Additive Explanations)~\citep{lundberg2017unified}.

Together, this SHAP analysis offers a complementary interpretability perspective—highlighting which features are critical for performance when using only handcrafted descriptors, and reinforcing the clinical plausibility of the vocal patterns captured by \SysName’s learned representations.

\begin{figure*}[htbp]
\centering
\includegraphics[width=\textwidth]{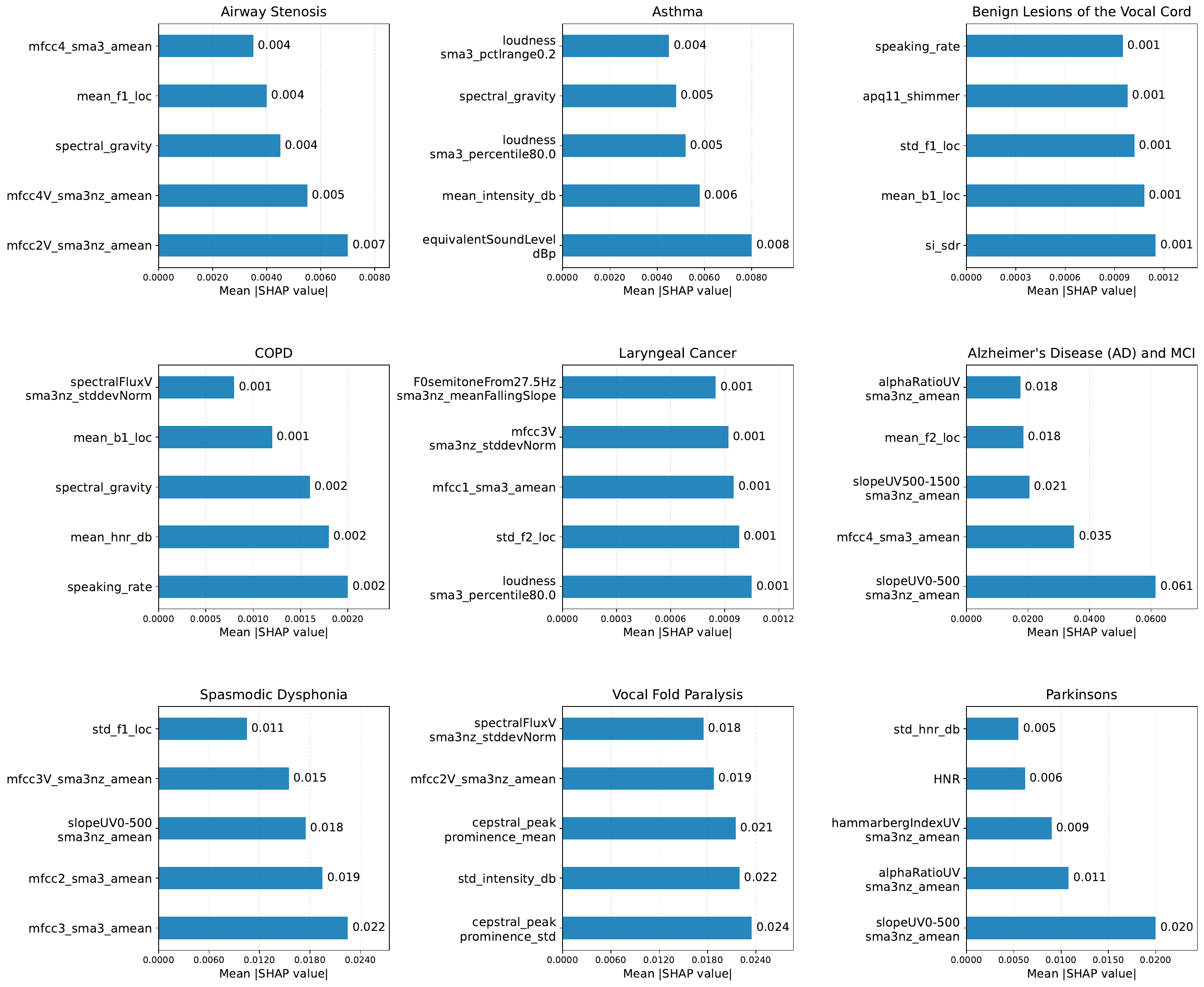}
\caption{SHAP-based acoustic feature attribution across nine disease classification tasks. Each subplot presents the top five features ranked by mean absolute SHAP value, which reflects the average magnitude of a feature’s contribution to model predictions. 
Each bar indicates the discriminative importance of acoustic features for identifying the target disorder.
}
\label{fig:shap_bar_all}
\end{figure*}

\subsection{Handcrafted Acoustic Feature Set and MLP Performance}
\label{sec:post_hoc_analysis_setup}
The handcrafted acoustic feature set in Bridge2AI-Voice dataset was compiled using established open-source signal processing toolkits, like openSMILE and Praat. This extensive pipeline extracts a diverse array of descriptors, encompassing spectral characteristics (e.g., spectral flux, roll-off), prosodic patterns (e.g., pitch-related features, speaking rate), voice quality measures (e.g., jitter, shimmer, Harmonic-to-Noise Ratio), and temporal dynamics. Each feature represents a global summary statistic computed over an entire speech recording, providing a structured, interpretable representation of vocal biomarkers.

As previously reported in Table~\ref{tab:highlevel-auroc} (Section~\ref{sec:comparison_baselines}), the MLP classifier trained solely on these handcrafted features demonstrated commendable performance across the high-level disorder categories. It achieved AUROC scores of $0.80 \pm 0.09$ for neurological disorders, $0.67 \pm 0.09$ for voice disorders, and $0.71 \pm 0.15$ for respiratory disorders. These results confirm that traditional, structured acoustic descriptors indeed capture discriminative patterns relevant to various health conditions, particularly for neurodegenerative disorders where subtle changes in motor control often manifest in measurable vocal characteristics, even without leveraging complex deep learning architectures or raw spectral inputs.

\subsection{Identifying Salient Acoustic Features with SHAP}
\label{sec:shap_analysis}

To understand which specific acoustic features drive the MLP's classification decisions for different disorders, we applied SHAP analysis. Figure~\ref{fig:shap_bar_all} illustrates the top-five most impactful features for each of the nine disease subtypes, ranked by their mean absolute SHAP value. This value indicates the average magnitude of a feature's contribution to the model's output across all samples. Each bar represents the relative discriminative importance of the feature for the corresponding disorder classification.

The analysis reveals distinct feature importance profiles across disorder categories, many of which align with known clinical voice characteristics. For \textit{neurological disorders} such as Parkinson's Disease and AD/MCI, features related to pitch stability and articulatory precision, including \texttt{slopeUV0-500} (reflecting pitch trajectory in unvoiced segments) and \texttt{alphaRatioUV} (related to spectral tilt), were prominent. The importance of \texttt{mean\_F2\_loc} (average second formant location) further suggests changes in vocal tract configuration, aligning with known deficits in motor control and prosodic modulation common in these conditions. In the case of \textit{respiratory disorders} like Airway Stenosis and COPD, features such as \texttt{spectral\_gravity} (spectral center of mass), MFCC-related statistics (e.g., \texttt{mfcc2V\_sma3nz}), and intensity measures (e.g., \texttt{mean\_intensity\_db}) proved significant. These often reflect changes in subglottal pressure, airflow turbulence, and overall vocal effort, consistent with impaired pulmonary function. Similarly, \textit{voice disorders} including Vocal Fold Paralysis and Spasmodic Dysphonia were primarily characterized by cepstral features (e.g., \texttt{cepstral\_peak\_prominence}) and other MFCC-derived metrics. These descriptors are sensitive to irregularities in glottal closure and vocal fold vibration patterns, which are hallmarks of laryngeal pathologies. Conversely, for conditions like Asthma where acoustic cues may be less consistently strong, SHAP values were generally more diffuse and lower in magnitude.

These findings collectively demonstrate that an interpretable model based on handcrafted acoustic features can not only achieve reasonable classification performance but also highlight pathophysiologically relevant vocal dimensions. This provides valuable context, suggesting that the acoustic cues exploited by our more complex model are likely rooted in these fundamental, clinically meaningful vocal characteristics.

\begin{table*}[t]
\centering
\caption{Cross-disease acoustic features that frequently appear in top-5 SHAP rankings. Avg SHAP denotes the mean absolute SHAP value of the feature, computed only from the diseases in which it was ranked among the top-5 most important features.}
\label{tab:universal_features}
\resizebox{0.95\textwidth}{!}{
\begin{tabular}{lcp{7.5cm}}
\toprule
\textbf{Feature} & \textbf{Avg SHAP} & \textbf{Associated Disorders} \\
\midrule
\texttt{slopeUV0-500\_\-sma3nz\_\-amean}  & 0.031 & Alzheimer\'s Disease (AD) and MCI, Spasmodic Dysphonia Laryngeal Tremor, Parkinson's \\
\texttt{mfcc2V\_\-sma3nz\_\-amean}        & 0.012 & Airway Stenosis, Spasmodic Dysphonia Laryngeal Tremor, Vocal Fold Paralysis \\
\texttt{spectral\_\-gravity}           & 0.005 & Airway Stenosis, Asthma, COPD \\
\texttt{loudness\_\-sma3\_\-percentile80.0}  & 0.004 & Airway Stenosis, Asthma, Benign Lesions of the Vocal Cord \\
\texttt{cepstral\_\-peak\_\-prominence\_\-mean}  & 0.017 & Vocal Fold Paralysis, Parkinson's \\
\texttt{mean\_\-intensity\_\-db}         & 0.004 & Laryngeal Cancer, Parkinson's \\
\texttt{mean\_\-b1\_\-loc}               & 0.003 & Asthma, Laryngeal Cancer \\
\bottomrule
\end{tabular}
}
\end{table*}

\vspace{0.3em}
\noindent
\textbf{Clinical Interpretation of Feature Relevance.}
Notably, the acoustic features identified as salient by SHAP correspond to well-established physiological correlates. Perturbations in \texttt{jitter} and \texttt{shimmer} reflect vocal fold instability and irregular phonation, commonly associated with laryngeal or neurogenic dysphonia \citep{lopes2017accuracy}. Another acoustic feature \texttt{Spectral\_gravity}, which measures energy distribution across frequencies, is indicative of airflow limitation and has been linked to compromised respiratory mechanics \citep{taylor2020age}. Cepstral features such as \texttt{cepstral\_peak\_prominence} relate to harmonic structure and are sensitive to glottal closure behavior, which is relevant in vocal fold paralysis and tremor \citep{balasubramanium2011cepstral}. Similarly, formant-based features like \texttt{mean\_F2\_loc} capture shifts in vocal tract resonance and articulatory posture, which are often impaired in neurodegenerative disorders \citep{mirarchi2017signal}. Collectively, these associations reinforce the biological validity of acoustic biomarkers and highlight their value as non-invasive indicators of clinical conditions, thereby supporting the foundational premise of voice-based disease modeling adopted in our primary framework.

\subsection{Universality versus Specificity of Acoustic Biomarkers}
\label{sec:universal_specific_features}

To investigate whether certain acoustic biomarkers are broadly indicative of vocal pathology (universal) versus specific to certain conditions, we analyzed the recurrence of the top-5 SHAP-ranked features across the nine disease-specific MLP models. Table~\ref{tab:universal_features} summarizes key features appearing in the top-5 list for multiple disorders, along with their average SHAP value (computed only for diseases where they ranked in the top-5) and associated conditions.

Several patterns emerge from this cross-condition analysis. Some features demonstrate broad relevance across related disorder families. For instance, \texttt{slopeUV0-500\_sma3nz\_amean} (average slope of unvoiced segments) shows high importance across three neurologically-related conditions (AD/MCI, Spasmodic Dysphonia, Parkinson's), suggesting its utility in capturing laryngeal motor control impairments common to these disorders. Similarly, \texttt{mfcc2V\_sma3nz\_amean} (second MFCC variant) is prominent across three structural/functional voice disorders (Airway Stenosis, Spasmodic Dysphonia, Vocal Fold Paralysis), likely reflecting alterations in vocal tract resonance due to anatomical or phonatory changes. Features like \texttt{spectral\_gravity} and \texttt{loudness\_sma3\_percentile80.0} also appear across three respiratory-related conditions, indicating systematic effects of respiratory limitations on sound energy and loudness.

This analysis reveals a hierarchy of acoustic biomarkers. Some features, like those related to unvoiced segment slopes or basic spectral balance, appear to be somewhat universal indicators of vocal system disruption across related disorder families. Others, while still shared between a couple of conditions, offer more targeted diagnostic information. This observation supports a layered approach to voice biomarker discovery, where both broadly indicative and highly specific features contribute to a comprehensive understanding of vocal health. The insights derived from these interpretable models reinforce the premise that speech contains rich, multi-faceted information about underlying health states, which more complex models like our \SysName~ framework can potentially exploit more effectively.

\end{document}